\documentclass[final]{svjour3}
\usepackage{graphicx}
\usepackage{rotating}
\usepackage{amssymb}
\usepackage{mathptmx}
\usepackage[numbers]{natbib}
\makeatletter
\journalname{Journal of Low Temperature Physics}

\bibpunct{}{}{,}{s}{}{,}
\usepackage{subfigure}
\usepackage{amsmath}
\usepackage{bm}
\usepackage{microtype}
\makeatletter
\journalname{Journal of Low Temperature Physics}

\def\Tr{$^3$H}
\def\mn{$m_\nu$}

\def\mum{$\mu$m}

\def\mus{$\mu$s}

\def\taur{$\tau_{R}$}
\def\tm{$t_M$}
\def\de{$\Delta E$}

\def\Aec{$A_\mathrm{EC}$}
\def\Qec{$Q_\mathrm{EC}$}
\def\fpp{$f_{pp}$}
\def\Nev{$N_{ev}$}
\def\ero{Er$_2$O$_3$}
\def\hoo{Ho$_2$O$_3$}
\def\Ho{$^{163}$Ho}
\def\Hom{$^{166m}$Ho}

\def\Er{$^{162}$Er}

\def\Ndet{$N_{det}$}
\def\mumux{$\mu$MUX}

\begin{document}

\newcommand{\hdblarrow}{H\makebox[0.9ex][l]{$\downdownarrows$}-}
\title{Status of the HOLMES experiment to directly measure the neutrino mass}

\author{A.~Nucciotti$^{4,5}$ \and B.~Alpert$^{1}$ 
\and M.~Balata$^{9}$
\and D.~Becker$^{1}$ \and D.~Bennett$^{1}$ 
\and A.~Bevilacqua$^{2,3}$
\and M.~Biasotti$^{2,3}$ 
\and V.~Ceriale$^{2,3}$ 
\and G.~Ceruti$^{5}$ 
\and D.~Corsini$^{2,3}$ 
\and M.~De~Gerone$^{2,3}$ \and R.~Dressler$^{6}$ \and M.~Faverzani$^{4,5}$ \and E.~Ferri$^{4,5}$ \and J.~Fowler$^{1}$ 
\and J.~Gard$^{1}$ \and F.~Gatti$^{2,3}$ \and A.~Giachero$^{5}$ \and J.~Hays-Wehle$^{1}$ \and S.~Heinitz$^{6}$ \and G.~Hilton$^{1}$ \and U.~K\"{o}ster$^{7}$ \and M.~Lusignoli$^{8}$ 
\and J.~Mates$^{1}$ \and S.~Nisi$^{9}$ 
\and  A.~Orlando$^{5}$ 
\and L.~Parodi$^{2,3}$ 
\and G.~Pessina$^{5}$ 
\and A.~Puiu$^{4,5}$ \and S.~Ragazzi$^{4,5}$ \and C.~Reintsema$^{1}$ \and M.~Ribeiro-Gomez$^{10}$ \and D.~Schmidt$^{1}$ \and D.~Schuman$^{1}$ \and F.~Siccardi$^{2,3}$ 
\and D.~Swetz$^{1}$ 
\and J.~Ullom$^{1}$ \and L.~Vale$^{1}$}

\institute{1: National Institute of Standards and Technology, Boulder, CO, USA\\
2: Dipartimento di Fisica, Universit\`{a} di Genova, Genova, Italy\\
3: Istituto Nazionale di Fisica Nucleare, Sezione di Genova, Genova, Italy\\
4: Dipartimento di Fisica, Universit\`{a} di Milano-Bicocca, Milano, Italy\\
5: Istituto Nazionale di Fisica Nucleare, Sezione di Milano-Bicocca, Milano, Italy\\
6: Paul Scherrer Institut, Villigen, Switzerland\\
7: Institut Laue-Langeving, Grenoble, France\\
8: Istituto Nazionale di Fisica Nucleare, Sezione di Roma 1, Roma, Italy\\
9: Laboratori Nazionali del Gran Sasso, INFN, Assergi (AQ), Italy\\
10: Multidisciplinary Centre for Astrophysics (CENTRA-IST), University of Lisbon, Lisbon, Portugal
}

\maketitle

\begin{abstract}

The assessment of neutrino absolute mass scale is still a crucial challenge in today particle physics and cosmology.  Beta or electron capture spectrum end-point study is currently the only experimental method which can provide a model independent measurement of the absolute scale of neutrino mass. HOLMES is an experiment funded by the European Research Council  to directly measure the neutrino mass. HOLMES will perform a calorimetric measurement of the energy released in the electron capture decay of the artificial isotope \Ho. 
In a calorimetric measurement the energy released in the decay process is entirely contained into the detector, except for the fraction taken away by the neutrino. This approach eliminates both the issues related to the use of an external source and the systematic uncertainties arising from decays on excited final states. The most suitable detectors for this type of measurement are low temperature thermal detectors, where all the energy released into an absorber is converted into a temperature increase that can be measured by a sensitive thermometer directly coupled with the absorber. This measurement was originally proposed in 1982 by A. De Rujula and M. Lusignoli, but only in the last decade the technological progress in detectors development has allowed to design a sensitive experiment.
HOLMES plans to deploy a large array of low temperature microcalorimeters with implanted \Ho\ nuclei. 
In this contribution we outline the HOLMES project with its physics reach and technical challenges, along with  its status and perspectives.  

\end{abstract}
\keywords{Neutrino mass measurement, Electron capture, Holmium, Transition Edge Sensors.}

\section{Introduction}
The HOLMES experiment aims at directly measuring the electron neutrino mass using the electron capture (EC) decay of \Ho\ [\citep{alpert15}].
HOLMES performs a calorimetric measurement of the energy released in the decay of \Ho\ to measure all the atomic de-excitation energy, except the fraction carried away by the neutrino [\citep{ltd-mnu}].  The direct measurement exploits only energy and momentum conservation and it is therefore completely model-independent. At the same time, the calorimetric measurement eliminates systematic uncertainties arising from the use of external beta sources, as in experiments with beta spectrometers, and minimizes the effect of the atomic de-excitation process uncertainties.
\begin{figure}[b]
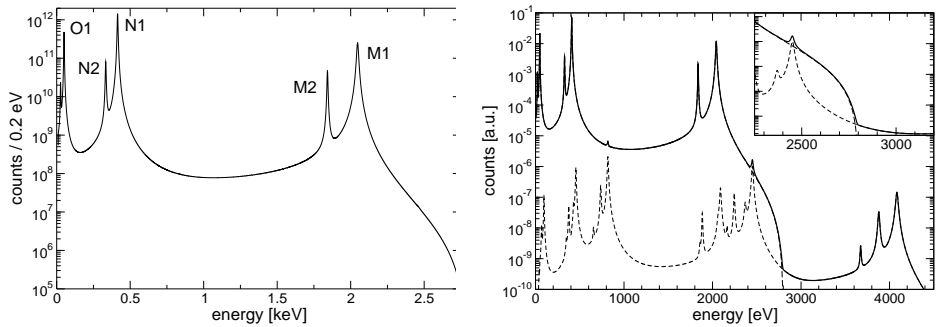
%
\begin{center}
\includegraphics[width=.49\textwidth]{olmio_cfr_Q-new_BW.eps}
\hfill
\includegraphics[width=.49\textwidth]{Ho-singlehole-pup_BW.eps}
\caption{\label{fig:calspe}\label{fig:ppspe} {\it Left.}\,Calculated \Ho\ EC calorimetric spectrum for $Q=2.8$\,keV, $\Delta E=2$\,eV, and $N_{ev}=10^{14}$. {\it Right.}\,Calculated experimental \Ho\ EC calorimetric spectrum for $Q=2.8$\,keV, $\Delta E=2$\,eV, $f_{pp}=10^{-4}$, and $N_{ev}=10^{14}$ (solid). The pile-up spectrum is the dashed curve .}
\end{center}
\end{figure}
The expected calorimetric spectrum is shown in fig.\ref{fig:calspe} (left): the distribution of the measured de-excitation energy, carried mostly by electrons with energies up to about 2\,keV, presents lines at the binding energies of the captured electrons. These lines have a natural width of a few eV, therefore the actual spectrum is a continuum with marked peaks with Breit-Wigner shapes. 
The spectral end-point is shaped by the same neutrino phase space factor $(Q_\mathrm{EC}-E)\sqrt{(Q_\mathrm{EC}-E)^2-m_{\nu}^2}$ that appears in a beta decay spectrum, where \Qec\ is the EC transition energy. A finite neutrino mass \mn\ causes a deformation of the energy spectrum which is truncated at \Qec-\mn.
The sensitivity of this approach depends on the nearness of \Qec\ to the binding energy of one of the captured electrons: \Ho\ was proposed in [\citep{lusignoli}] as an ideal isotope with a \Qec\ between 2.5\,keV and 3.0\,keV -- with 2.55\,keV as raccomended value [\citep{reich2010nuclear}],-- close to the binding energy of about 2.0\,keV for the M1 electrons. Recent measurements [\citep{eliseev}] established that \Qec\ is $2833\pm30_{stat}\pm15_{sys}$\,eV, therefore not as close to the M1 shell binding energy as initially hoped for. 

The spectrum shown in fig.\,\ref{fig:calspe} (left) is calculated neglecting the second order effects in the atomic de-excitation cascade -- shake up and shake off -- which have been recently considered in many publications and that are summarized in [\citep{ltd-mnu}] and references therein. 

\section{The HOLMES experiment design}
\begin{figure}[t!]
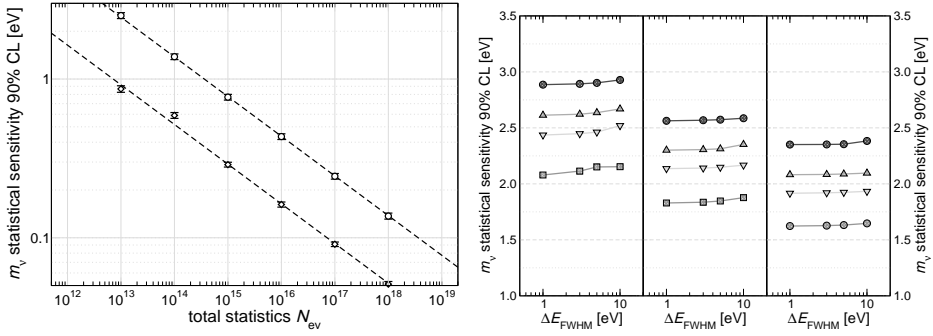
%
\begin{center}
\includegraphics[width=.49\textwidth]{q28_new_BW.eps}
\hfill
\includegraphics[width=.49\textwidth]{hosim10.eps}
\caption{\protect\label{fig:q28}\label{fig:hosim10}{\it Left.} Monte Carlo estimated statistical sensitivity for $\Delta E=1$\,eV, $\tau_R=1$\,\mus, and for both $f_{pp}=10^{-3}$ and $10^{-6}$ (from top to bottom). The dashed lines are the extrapolated curves using a $N_{ev}^{-1/4}$ scaling law. {\it Right.}\,Monte Carlo estimated statistical sensitivity for a fixed exposure $t_M \times N_{det}$ of about 3000 detector$\times$year and for different \Aec\ (from left to right panel: 30, 100, and 300\,Bq/detector) and   $\tau_R$ (from top to bottom: 10, 5, 3, and 1\,\mus)}
\end{center}
\end{figure}

The sensitivity of calorimetric experiments is affected by a major drawback of this approach. In a calorimeter the whole decay spectrum is acquired and the decaying isotope activity \Aec\ must be\ restrained to limit the rate of accidental coincidences. Time unresolved coincidences (pile-up) show up in the spectrum as an additional background which worsens the statistical sensitivity. The right side of fig\,\ref{fig:ppspe} shows the pile-up spectrum whose normalization \fpp\ relative to single events spectrum (fig.\,\ref{fig:calspe}, left) is $f_{pp}\approx \tau_R A_\mathrm{EC}$, where \taur\ is the time resolution of the detector.
Therefore, when planning a calorimetric experiment, a trade off must be found between the quest for collecting a large statistics in a short time and the need for limiting the pile-up spectrum amplitude. For fixed measuring time \tm\ and detector performance -- \de\ and \taur, i.e. energy and time resolution, respectively, -- the experimental parameters to tweak are the number of detectors \Ndet\ and the isotope activity \Aec\ in each of them.
The statistical sensitivity as a function of the above experimental parameters can be investigated exploiting the Monte Carlo methods described in [\citep{nucciotti_ECsens}]. 
The left side of fig.\,\ref{fig:q28} shows that the statistical sensitivity improves with the total statistics \Nev\ as $1/\sqrt[4]N_{ev}$ and that reaching a sensitivity below 2\,eV requires around $10^{13}$ decays. While matching the current 2\,eV sensitivity obtained by spectrometers with \Tr\ [\citep{lobashev,kraus}] is a meaningful first target for current \Ho\ experiments, the goal for future experiments must be at least 0.1\,eV -- i.e. better than the KATRIN experiment goal [\citep{weinheimer}]. 
From the left side of fig.\,\ref{fig:q28} it is apparent how the 0.1\,eV target calls for statistics larger than about $10^{17}$ events to be collected by arrays of more than $10^7$ pixels.

HOLMES optimal configuration is found by means of the  simulations plotted in the right side of fig.\,\ref{fig:hosim10} and by appreciating that, for fixed exposure $t_M \times N_{det}$ and time resolution \taur, the statistical sensitivity constantly improves for increasing pixel activities \Aec.
As a consequence, in spite of the increasing fraction of pile-up events \fpp, it is winning to have the largest tolerable pixel activity. With this in mind, HOLMES plans to deploy an array with 1024 pixels, each with an \Ho\ activity \Aec\ of about 300\,Bq. The right panel of fig.\,\ref{fig:hosim10} shows that to reach a statistical sensitivity below 2\,eV the time resolution \taur\ plays the main role, while the energy resolution \de\ could be anywhere below about 10\,eV.  
In particular \taur\ must be better than about 3\,\mus, which translates in a fraction of pile-up \fpp\ better than about $10^{-3}$.

Because of the very low fraction of decays in the region of interest close to \Qec, the background  is another factor impacting the statistical sensitivity of end-point neutrino mass measurements.
A constant background $b$ is negligible as long as it is much smaller
than the pile-up spectrum, that is when $b \ll \approx A_\mathrm{EC}f_{pp}/2Q_\mathrm{EC}$, as it is confirmed also by Monte Carlo simulations [\citep{nucciotti_ECsens}]. 
Therefore, a large activity \Aec\ with a correspondingly large pile-up fraction \fpp\ makes experiments relatively insensitive to cosmic rays and to environmental radioactivity. 
For HOLMES this translates in the requirement that the background level at the end-point must be lower than about 0.1\,count/eV/day/det.
From Geant4 Monte Carlo simulations and comparison with past experiments, HOLMES' background due external sources is expected to be of the order of $10^{-4}$\,count/eV/day/det at sea-level.
Indeed HOLMES' major source of background is the internally present $^{166m}$Ho, an isotope which is produced along with \Ho\ [\citep{Enge12}].
This $\beta$ decaying isotope, with a half life of about 1200\,years and a $Q$-value of about $1854$\,keV, according to Geant4 Monte Carlo simulations causes a background below 5\,keV of about 0.5\,count/eV/day/det/Bq(\Hom). The \Hom\ activity in the pixels must be then kept lower than 0.2\,Bq for a \Ho\ activity \Aec\ of 300\,Bq, or a factor 1500 lower, i.e. the number of contaminating \Hom\ nuclei must be about a factor 6000 lower than that of \Ho.  

As discussed in more details in the following, the HOLMES project is divided in tasks which are carried out in parallel and on which the HOLMES groups
are making steady progresses [\citep{giachero}]. The tasks are 1) the \Ho\ isotope production and purification, 2) the isotope embedding, 3) the TES pixel and array design and testing, 4) the microwave multiplexed read-out system, and 5) the Digital Acquisition system (DAQ). 

\section{$^{163}$Ho production and embedding}
The \Ho\ isotope needed to carry out the experiment is produced at the nuclear reactor sited at the Institute Laue-Langevin (ILL, Grenoble, France) by irradiating \ero\ samples  enriched in \Er\ in a thermal neutron flux of about 10$^{15}$\,n/s/cm$^2$. 
The \ero\ samples are purified before irradiation and the accumulated holmium is radiochemically separated in hot-cells after the neutron irradiation. Both processes have been developed and optimized at the Paul Scherrer Institute (PSI, Zurich, CH).
Two samples of enriched \ero\ have been already irradiated at ILL and processed at PSI. Inductive Coupled Plasma Mass Spectroscopy (ICPMS) analysis performed at LNGS and PSI on the two samples before and after chemical purification at PSI demonstrated a production of a sufficiently radiopure sample of about 43\,MBq of \Ho\ (along with about 50\,kBq of \Hom, measured by $\gamma$-spectroscopy). 
The analysis of these samples yielded also a preliminary estimate of the \Ho\ production yield including all these steps.
We estimate that the whole HOLMES program requires a total amount of about 250\,MBq of \Ho. This estimate is obtained considering all the processing efficiencies involved from isotope production to embedding in the detectors: these include the above mentioned yield, but also the efficiencies related to the ion implantation process. Since not all the efficiencies are known yet, for the above estimate we made conservative guesses to come up with a global efficiency of the order of 0.1\%.    
About 540\,mg of \ero\ enriched in $^{162}$Er at 25\% were irradiated at ILL for about 50\,days until early 2017. Along with this sample we also re-irradiated about 100\,mg of \ero\ at 26.9\% which were left after the holmium chemical separation at PSI from last irradiated sample. 
We estimate a production of about 150\,MBq of \Ho\ which are expected to be enough both for testing the isotope embedding and for the production of the first 512 detectors. 

\begin{figure}[t!]%
\begin{center}
\includegraphics[width=.8\textwidth]{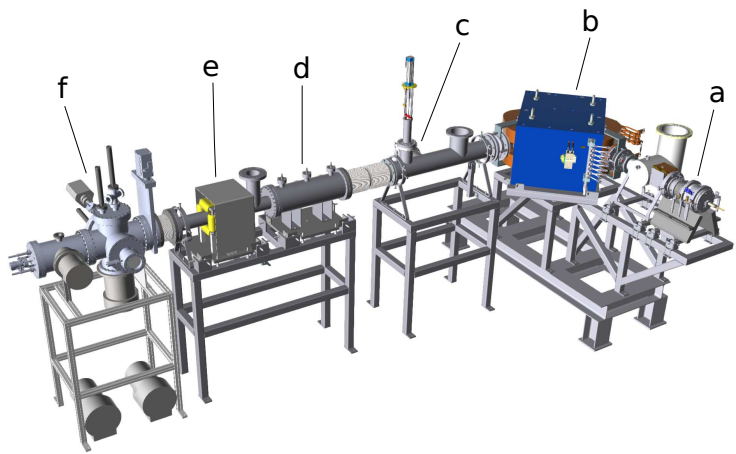}
\caption{\label{fig:embedding}3D model of the embedding system showing: (a) Penning sputter ion source, (b) analyzing magnet, (c) analyzing slit, (d) electrostatic triplet focusing stage, (e) XY magnetic beam scanner, and (f) Target Chamber. (Color figure on-line)}
\end{center}
\end{figure}

\Ho\ is introduced in the absorbers of the HOLMES' low temperature microcalorimeters by means of
custom system which combines a high efficiency Penning sputter ion source, a mass analyzing magnet, an electrostatic triplet focusing stage, and an XY magnetic beam scanner. The system is shown in fig.\ref{fig:embedding} and it is described in more details in [\citep{gallucci}]. 
The system is designed to achieve an optimal mass separation for \Ho, thereby eliminating  
other trace contaminants not removed by chemical methods at PSI, such as $^{166m}$Ho. The magnetic selection also minimizes the implantation of other stable isotopes such as $^{165}$Ho, which would only increase the detector heat capacity, and prevents the implantation of holmium in chemical forms other than metallic, which would likely cause chemical shifts of the end-point energy.
The embedding system includes a focusing stage designed to have a beam cross-section on the detectors of about 4\,mm thereby maximizing the geometrical efficiency of the ion implantation (see [\citep{orlando}]). During ion implantation, the detectors are hosted in the UHV Target Chamber which is equipped with an ion beam assisted sputtering system to control the \Ho\ concentration in the detector absorbers, to compensate the absorber atom sputtering caused by the ion implantation, and to deposit the final absorber layer to complete the \Ho\ embedding [\citep{orlando}].
The metallic cathode for the ion source is made out of an inter-metallic alloy containing \Ho\ in a metallic form. The metallic \Ho\ is produced in an evaporation chamber by thermal reduction and distillation of the \hoo\ separated at PSI after neutron irradiation. The details of the ion source cathode fabrication are described in [\citep{gallucci}]. 
The whole embedding system is expected to be ready for detector implantation testing early in 2018.

\section{TES microcalorimeters, microwave multiplexed read-out, and DAQ}
The detectors used for the HOLMES experiment are Mo/Cu TES on Si$_2$N$_3$ membrane with 2\,\mum\, thick gold absorbers. The pixel design has been optimized to match the experimental specifications in terms of energy and time resolution, pulse duration, and \Ho\ decay radiation full absorption [\citep{jimltd,faverzani}].
As explained above the most critical parameter is the time resolution. By means of Monte Carlo simulations we find that with rise and decay time constants of about 10\,\mus\ and 100\,\mus, respectively, and with signal sampling rate of at least 500\,kHz it is possible to obtain a time resolution better than 3\,\mus\ exploiting discrimination algorithms based on Singular Value Decomposition [\citep{alpertltd}] or Wiener filtering [\citep{ferri}]. 

The HOLMES arrays are read-out with the microwave multiplexing ($\mu$MUX) developed by NIST, which is based on  rf-SQUIDs as input devices  [\citep{mates}]. 
The $\mu$MUX read-out leverages    the Software Defined Radio (SDR) implemented in the firmware of a ROACH-2 board by NIST [\citep{decker}].
The HOLMES DAQ system is presently composed by one ROACH-2 system with ADC (550\,MS/s, 12\,bit, 2 channels), DAC (for comb generation), and IF circuitry (for signal up- and down-conversion) realized with discrete components (IQ mixers, power dividers, attenuators, and RF amplifiers).
We used this set-up to characterize two $\mu$MUX chips from NIST (MUX16a and MUX17a) [\citep{puiu}]. The two \mumux\ chips provide 33  2\,MHz wide resonances spaced by 14\,MHz in a 500\,MHz band.
The multiplexing factor $n_\mathrm{TES}$ achievable with one ROACH-2 board is $n_\mathrm{TES} \approx 0.005 f_\mathrm{ADC} \tau_{rise}$ for a signal sampling frequency of about 0.5\,MHz, where $f_\mathrm{ADC}$ is the SDR ADC sampling frequency of 550 MHz. Considering the design $\tau_{rise}$ of about 10\,\mus, the multiplexing factor for the HOLMES DAQ can be as high as 32 and therefore to read-out the full 1024 pixel array a total of 32 ROACH-2 systems is required.

Two $6\times4$ prototype arrays  were fabricated at  NIST (Boulder, Co, USA) with slight variations in the TES pixel designs and were characterized using both $^{55}$Fe and a fluorescence multi-line source. The prototype pixel signals were read-out either with a 2-channel homodyne circuit [\citep{ferrielba,puiultd}] or with the 16-channel ROACH-2 DAQ. The results of these measurements, which are described in details in [\citep{puiu}], confirm that the baseline pixel design, along with a proper tuning of the bias network inductance $L$, while matching the desired time constants, provide an energy resolution well below 10\,eV FWHM at \Qec. The measurements carried out using the ROACH-2 DAQ also proves that the HOLMES DAQ is ready for the experiment.

According to these results, we have now finalized the design of the $4\times16$ sub-arrays which will be deployed for the HOLMES experiment [\citep{orlando}]: the design aims to 1) minimize the signal bandwidth limitations due to stray self-inductance of the read-out leads,  2) minimize the signal cross-talk due to mutual inductance between read-out lines, and 3) maximize the geometrical filling for an optimal implantation efficiency.
The arrays are fabricated in a two step process [\citep{orlando}]. 
The arrays are provided by NIST with a 1\,\mum\ gold layer and they are further processed after ion-implantation: first in the Target Chamber the thin (few 100\AA{}) layer of Au:\Ho\ is covered by a second 1\,\mum\ gold layer to fully encapsulate the \Ho\ source, then the Si$_2$N$_3$  membranes are released by means of a Deep Reactive Ion Etching. Preliminary tests on dummy samples from NIST are in progress to tune the DRIE process and define the details of this two-step fabrication process [\citep{orlando}]. 

\section{Future plans}
With the embedding system fully operational early in 2018, HOLMES will start soon to optimize the isotope implantation process along with the two-step array fabrication process. 
As soon as the first implanted arrays will be available, the first high statistics calorimetric measurements of the \Ho\ decay will provide relevant information on the spectral shape and new competitive limit on \mn.  

\begin{acknowledgements}
The HOLMES experiment is funded by the European Research Council under the European Union Seventh Framework Programme (FP7/2007-2013)/ERC Grant Agreement no. 340321. We also acknowledge support from INFN for the MARE project, from the NIST Innovations in Measurement Science program for the TES detector development, and from Funda\c{c}\~{a}o para a Ci\^{e}ncia e a Tecnologia (PTDC/FIS/116719/2010) for providing the enriched Er$_2$O$_3$ used in preliminary $^{163}$Ho production by means of neutron irradiation. 
\end{acknowledgements}

\end{document}